\documentclass[10pt]{article}

\usepackage{amsmath}

\usepackage{array}

\usepackage{appendix}

\usepackage{tocloft}                   

\usepackage{graphicx}

\usepackage{amsfonts}

\usepackage{amssymb}

\usepackage{mathrsfs}

\usepackage{yfonts}

\usepackage{euscript}

\usepackage{centernot}                 

\usepackage{ifsym}                     

\usepackage{upgreek}

\usepackage{mathtools}

\usepackage{bm}

\usepackage{color}

\usepackage{slantsc}
\usepackage{calligra}

\usepackage{bbold}          

\usepackage[T1]{fontenc}

\usepackage{epsf}

\usepackage{latexsym}

\usepackage{tipa}

\usepackage{makeidx}

\makeindex

\def\be{\begin{equation}}

\def\ee{\end{equation}}

\def\beq{\begin{equation}}

\def\eeq{\end{equation}}

\def\bea{\begin{eqnarray}}

\def\eea{\end{eqnarray}}

\def\!{\hspace{-1.6667em}}

\def\m{\mbox{ }}

\def\!{\hspace{-1.6667em}}

\def\n{\noindent}

\def\u{\underline}

\def\es{\m = \m}

\def\:={\m := \m}

\def\=:{\m =: \m}

\def\sbig{\mbox{\scriptsize\boldmath$g$}}

\def\biO{\mbox{\boldmath$O$}}

\def\mJ{\mbox{J}}  

\def\mK{\mbox{K}}

\def\me{\mbox{e}}

\def\mh{\mbox{h}}

\def\mi{\mbox{i}}

\def\ml{\mbox{l}}

\def\mn{\mbox{n}}  

\def\mo{\mbox{o}}

\def\mr{\mbox{r}}

\def\ms{\mbox{s}}

\def\bupSigma{\mbox{\boldmath$\Sigma$}}                 
\def\sbupSigma{\mbox{\scriptsize\boldmath$\Sigma$}}     

\def\se{\mbox{\scriptsize e}}

\def\si{\mbox{\scriptsize i}}

\def\sm{\mbox{\scriptsize m}}

\def\sn{\mbox{\scriptsize n}}

\def\sss{\mbox{\scriptsize s}}  

\def\sv{\mbox{\scriptsize v}}

\def\sG{\mbox{\scriptsize G}}

\def\sR{\mbox{\scriptsize R}}

\def\sumi2{\sum\mbox{}_{\mbox{}_{\mbox{\scriptsize $i$=1}}}^2}

\def\sumi3{\sum\mbox{}_{\mbox{}_{\mbox{\scriptsize $i$=1}}}^3}

\def\sumABcycles3{\sum\mbox{}_{\mbox{}_{\mbox{\scriptsize cycles $A,B$=1}}}^{3}}

\def\sumCDcycles3{\sum\mbox{}_{\mbox{}_{\mbox{\scriptsize cycles $C,D$=1}}}^{3}}

\def\sumj3{\sum\mbox{}_{\mbox{}_{\mbox{\scriptsize $j$=1}}}^3}

\def\sumk3{\sum\mbox{}_{\mbox{}_{\mbox{\scriptsize $k$=1}}}^3}

\def\prodiA1{\prod\mbox{}_{\mbox{}_{\mbox{\scriptsize $i$=1}}}^{A - 1}}

\def\d{\textrm{d}}                                                  

\def\pa{\partial}                                                   

\def\lFrg{\mbox{\Large$\mathfrak{g}$}}                         
\def\nFrg{\mbox{\large$\mathfrak{g}$}}                         

\def\Hilb{\mbox{{\boldmath$\mathfrak{H}$}ilb}}                 
\def\FrQ{\mbox{\Large $\mathfrak{q}$}}                               
\def\sFrQ{\mbox{\large $\mathfrak{q}$}}                              
\def\Phase{\mbox{{\boldmath$\mathfrak{P}$}hase}}                     

\def\bFrR{\mbox{\boldmath$\mathfrak{R}$}}                            
\def\Rig-Phase{\bFrR\mbox{ig-}\Phase}                                
                                                                													   
\def\bFrM{\mbox{\boldmath${\mathfrak{M}}$}}                             

\def\Positive-Modespace{\mbox{{\boldmath$\mathfrak{M}$}odespace$^+$}}

\def\POSITIVE-MODESPACE{\mbox{{\boldmath$\mathfrak{M}$}ODESPACE$^+$}}
                                                                                                                             														
\def\Riem{\bFrR\mbox{iem}}                                           

\def\Kin-Hilb{\mbox{{\boldmath$\mathfrak{K}$}in-\Hilb}}                     

\def\Mid-Hilb{\mbox{{\boldmath$\mathfrak{M}$}id-\Hilb}}                     

\def\Dyn-Hilb{\mbox{{\boldmath$\mathfrak{D}$}yn-\Hilb}}                     

\def\5Star{\mbox{\Large$\star$}}              

\begin{document}

\begin{center}

\vspace{0.1in}

\Large{\bf A Local Resolution of the Problem of Time} \normalsize

\vspace{0.1in}

{\large \bf Edward Anderson$^*$}

\vspace{.05in}

\end{center}

\begin{abstract}

We here announce and outline a solution of this major and longstanding foundational problem, dealing with all seven of its heavily-interrelated local facets.

\end{abstract}

\m 

\n PACS: 04.20.Fy and Cv. 

\vspace{0.1in}
  
\n $^*$ Dr.E.Anderson.Maths.Physics@protonmail.com . 
Work done at {\sl Queen Mary, University of London; 
                  Peterhouse Cambridge; 
				  IFT Universidad Autonoma de Madrid; 
				  DAMTP, Cambridge; 
				  Universit\'{e} Paris VII}.

\section{Introduction}

Let us assume familiarity with the Isham--Kucha\v{r} conceptual classification of the Problem of Time into 8 facets (see \cite{Kuchar92, I93} or \cite{APoT} for a summary), 
which grew from previous observations by Dirac \cite{DiracObs, Dirac}, Wheeler \cite{Battelle} and DeWitt \cite{DeWitt67}.    
We update and generalize many of these, renaming them in the process, as per Fig 1. 
The point of the current article is a {\sl brief} account; the details are already available in the more extensive references indicated.
%
{            \begin{figure}[!ht]
\centering
\includegraphics[width=0.75\textwidth]{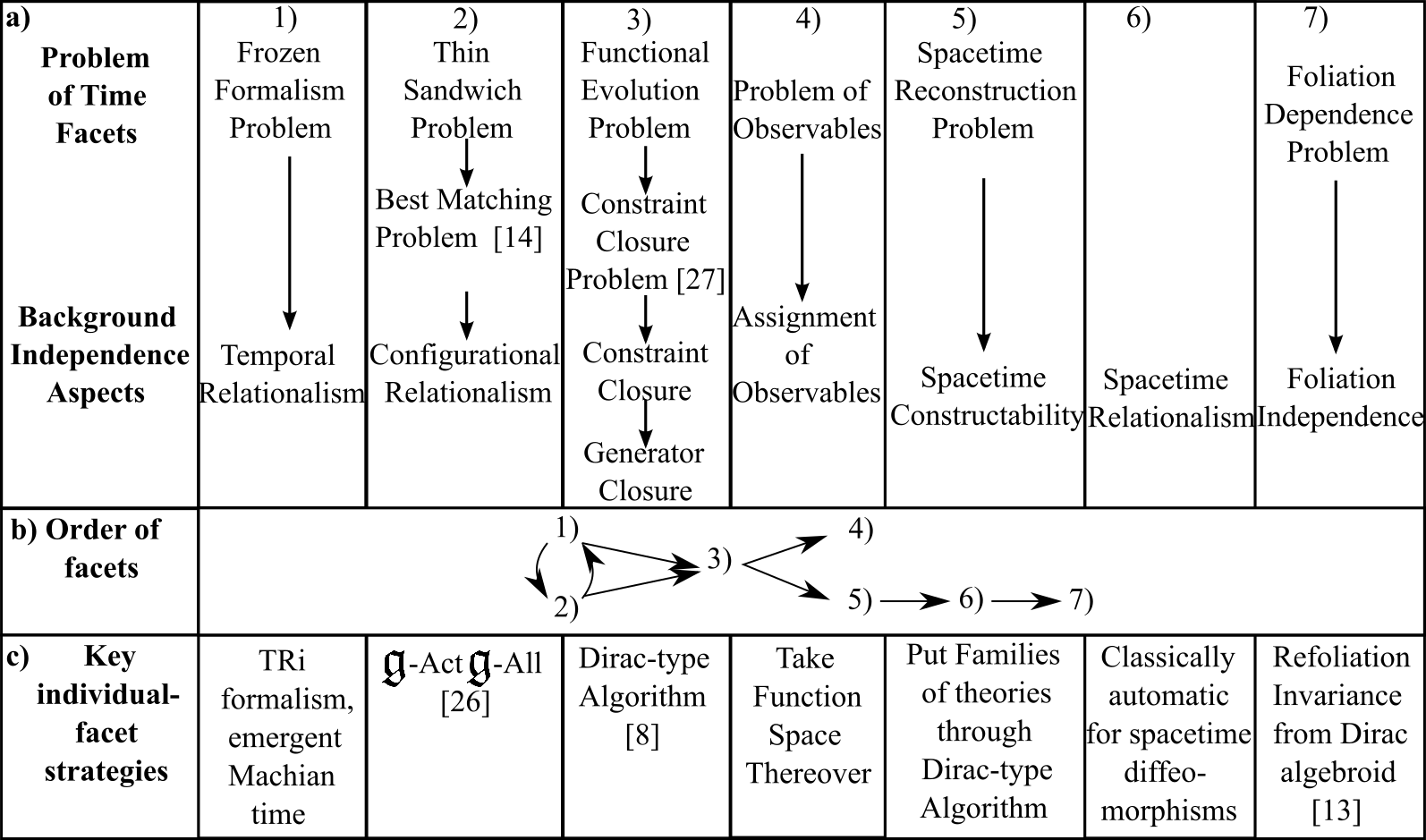}
\caption[Text der im Bilderverzeichnis auftaucht]{\footnotesize{a) Evolution of conceptualization and nomenclature of Problem of Time facets. 
The first row are Kucha\v{r} and Isham's \cite{Kuchar92, I93}. 
This Figure moreover links these facet names and concepts to Background Independence \cite{A64-67, Giu06} aspects \cite{APoT3, ABook}, 
from the position that difficulties with implementing Background Independence aspects {\sl result in} Problem of Time facets.
All bar the last of these aspects are already classically present. 
b)  The order in which the facets are incorporated, itself resolving a longstanding problem \cite{Kuchar92, I93, Kuchar93}.
c) The key strategies used in the current Article's resolution.                                             } }
\label{Evol-Fac}\end{figure}            }

\section{Constraint Providers}

As regards {\it Temporal Relationalism} \cite{FileR}, we do not just incorporate its primary-level Leibnizian timelessness for the universe as a whole \cite{L} 
via a {\it Temporal Relationalism implementing} (TRi) action \cite{B94I}, most commonly of the homogeneous quadratic `Jacobi action' form\footnote{Here 
$\d s = \sqrt{\d Q \cdot M \cdot \d Q}$ is the kinetic line element, for $\d Q$ changes in configuration $Q$, which in turn form an incipient configuration space $\sFrQ$.  
Also $W$ is a `potential factor': $E - V$ for Mechanics, for total energy $E$ and potential energy $V$, 
or $R - 2 \, \Lambda$ for GR, whose incipient configurations are metrics $h_{ab}$ on 3-space $\sbupSigma$, 
with $R$ the corresponding spatial Ricci scalar and $\Lambda$ the cosmological constant.} 

\n \be 
S  \es  \int \d s \sqrt{2 \, W}  \m .  
\label{S}
\ee 
We rather incorporate Leibnizian timelessness a fortiori by using an entire {\it TRiPoD} (Principles of Dynamics \cite{TRiPoD}), and more, as per Secs 5 and 6.  
By this greater endeavour, we manage to {\sl remain entirely} within Temporal Relationalism as we handle each subsequent facet in turn.

\m 

\n TRi actions moreover provide {\it primary constraints}, by a basic argument of Dirac \cite{Dirac}.  
The `energy constraint' ${\cal E}$ of Mechanics \cite{FileR} and the Hamiltonian constraint ${\cal H}$ of GR \cite{ADM} can be taken to arise in this manner, 
as constraints quadratic in their momenta $P$  (conjugate to $Q$)

\n \be 
{\cal C}\mh\mr\mo\mn\mo\ms  \:=  P \cdot N \cdot P \, / \, 2 - W \es 0 
\ee 
for $N$ the inverse of $\d s$'s metric $M$.  
\n ${\cal C}\mh\mr\mo\mn\mo\ms$ receives moreover the interpretation of an `equation of time', being rearrangeable to 

\n \be
t^{\se\sm}  \es  \int \d s/\sqrt{{2 \, W}}  \m : 
\ee
an emergent Machian \cite{M} statement that {\sl time is to be abstracted from change} ($\d s$ being a function of $\d Q$).
While any \cite{R04} and all \cite{B94I} change have been used elsewhere, 
what a detailed analysis reveals to actually be required (Chapter 15 of \cite{ABook}) is a {\it sufficient totality of locally relevant change} (STLRC). 
The time abstracted from this is {\it generalized local ephemeris time} (GLET); this is even {\sl even more like} ephemeris time \cite{Clemence} that \cite{B94I}'s notion of time.

\m 

\n An incipient approach to {\it Configurational Relationalism} \cite{FileR} is to correct actions' velocities additively with multipliers \cite{BB82} 
corresponding to a group $\lFrg$ of physically-irrelevant motions.   
While this ruins the TRi homogenity of the action $S$ -- the first of many facet interferences -- one can instead correct velocities by auxiliary cyclic velocities, 
or, better, correct changes by auxiliary cyclic changes $\d \alpha$:  

\n \be 
\d Q \m \longrightarrow \m \d Q - \pounds_{\d \alpha} Q   \m .
\ee
for $\pounds$ the Lie derivative that implements $\lFrg$ corrections.
Using this means of correction, we succeed in jointly incorporating Temporal and Configurational Relationalism.  
Variation with respect to $\alpha$ moreover provides first-class constraints that are linear in their momenta, ${\cal F}\ml\mi\mn$; 
taking these into account sends one to the quotient configuration space $\FrQ/\lFrg$.
In the case of GR, $\FrQ = \Riem(\bupSigma)$:                       the space of $h_{ab}$, 
                   $\lFrg = Diff(\bupSigma)$                       (3-diffeomorphisms) produces the momentum constraint ${\cal M}_i$ \cite{ADM, Dirac} as per above, 
               and $\FrQ/\lFrg = \Riem(\bupSigma)/Diff(\bupSigma)$ is Wheeler's superspace \cite{Battelle} .  

\m 
 
\n In fact, resolving Configurational Relationalism requires us to correct more than just actions (distances, quantum operators...). 
The general approach is to {\sl act} with $\lFrg$ and then to perform an operation involving {\sl all} of $\lFrg$ \cite{FileR, ABook}, 

\n \be 
\biO  \m \longrightarrow \m\biO_{\nFrg\mbox{-}\si\sn\sv}  \:=  \mbox{\Large S}_{\sbig \, \in \,  \nFrg}  \stackrel{\rightarrow}{\lFrg}_{\sbig} \biO \mbox{ } .  
\ee 
This produces $\lFrg$-invariant versions of whatever objects $\biO$, as is most clear from the group-averaging procedure of basic Group and Representation Theory. 
The version for actions has an extremizing rather than averaging `all' operation. 
Note finally that we need to {\sl iterate} implementing Temporal and Configurational Relationalisms until consistency is attained.  
E.g.\ full GR's kinetic line element is 

\n \be 
\d s_{\sG\sR}                   \:=  M^{abcd}\{\d  - \pounds_{\d \u{F}}\}h_{ab}\{\d - \pounds_{\d \u{F}}\}h_{cd}   
\m  \mbox{ for } \m M^{abcd}  \:=  \sqrt{h}\{h^{ac}h^{bd} - h^{ab}h^{cd}\}
\ee   
the inverse of the DeWitt supermetric $N_{abcd}$, as occurs in th explicit form of the GR Hamiltonian constraint, 

\n \be 
{\cal H}  \:=  N_{abcd}p^{ab}p^{cd} + \sqrt{h} \{ 2 \, \Lambda - R \} \es 0  \m , 
\ee 
for gravitational momenta $p^{ab}$ conjugate to $h_{ab}$, and $\d \u{F}$ the TRi analogue of the ADM shift \cite{ADM}.  
Another interference is that finding an explicit $t^{\se\sm}$ in the presence of corrections requires prior elimination of $\d \alpha$ by use of ${\cal F}\ml\mi\mn$ \cite{ABook}.

\section{Constraint Closure}

We moreover need to assess whether the constraints provided by Temporal and Configurational Relationalism are consistent.  
The {\it Dirac Algorithm} \cite{Dirac, HT92} is a powerful Hamiltonian-level tool for assessing whether a set of constraints in hand is, or can be made, consistent. 
Its end product is a {\it constraint algebraic structure} that is closed under Poisson brackets.  
While the Dirac Algorithm is not itself Temporally Relational, it admits a TRi version: the {\it differential-almost-Dirac Algorithm} \cite{ABook}. 
This lies within the {\it differential almost-Hamiltonian formulation} \cite{TRiPoD, AM13, ABook}, admitting auxiliary cyclic differentials $\d \alpha$ as well as physical momenta, 
$\d\alpha$ being used to append constraints in place of the original Dirac Algorithm's use of multipliers. 

\m 

\n This program's main examples are 1) minisuperspace (spatially homogeneous GR), which has a single finite constraint, and so closes trivially. 
2) Relational Mechanics \cite{BB82, Kendall, FileR, AMech} and 3) Perturbatively Inhomogeneous GR \cite{HallHaw} both close \cite{ABook}. 
4) Full GR itself closes \cite{DiracAlg1, DiracAlg2, Tei73}, in the form of the {\it Dirac Algebroid}, schematically 

\n \be
\mbox{\bf \{} ( {\cal M}_i    \, | \,    \d L^i    ) \mbox{\bf ,} \, ( {\cal M}_j    \, | \,    \d M^j    ) \mbox{\bf \}}  \es  
              ( {\cal M}_i    \, | \, \, [ \d L, \d M ]^i )                                                                      \m ,
\label{Mom,Mom}
\ee

\n \be
\mbox{\bf \{} (    {\cal H}    \, | \,    \d J    ) \mbox{\bf ,} \, (    {\cal M}_i  \, | \,    \d L^i    ) \mbox{\bf \}}  \es  
              (    \pounds_{\d\underline{L}} {\cal H}    \, | \,    \d J    )                                                    \m , 
\label{Ham,Mom}
\ee

\n \be 
\mbox{\bf \{} (    {\cal H}    \, | \,    \d J    ) \mbox{\bf ,} \,(    {\cal H}    \, | \,    \d K    )\mbox{\bf \}}      \es  
              (    {\cal M}_i h^{ij}   \, | \,    \d J \, \overleftrightarrow{\pa}_j \d K    )                                 \m , 
\label{Ham,Ham}
\ee  
for $( \m | \m )$ the integral-over-space functional inner product, Lie bracket $[ \m , \m ]$ and TRi smearing functions $\d J, \d K, \d L^i$ and $\d M^i$.   
The third equation's right-hand-side containing the inverse of the spatial 3-metric $h^{ij}$ -- a function -- 
is why this is an {\sl algebroid} \cite{Bojowald} rather than an algebra; 
its containing ${\cal M}_i$ as well establishes ${\cal M}_i$ to be an {\sl integrability} of ${\cal H}$ \cite{MT72}.
By this, GR's ${\cal H}$ (and underlying Temporal Relationalism) cannot be entertained without ${\cal M}_i$ (or its underlying Configurational Relationalism).   
Relational Mechanics is moreover even less coupled \cite{ABook}), by which Temporal and Configurational Relationalism can be entertained piecemeal there.

\section{Observables}

\n The most primary notion of observables involves {\it Taking a Function Space Thereover}, referring in the classical canonical case to over phase space.  

\m 

\n Facet interference with the presence of constraints imposes the {\it commutation condition}  

\n \be 
\mbox{\bf \{}{\cal C}_A\mbox{\bf ,} \, B \mbox{\bf \}}  \es  0
\label{CO}
\ee 
on these observables functions, $B$. 
For this to be consistent \cite{ABeables}, ${\cal C}_A$ must moreover be a closed (sub)algebraic structure of constraints, 
by the ${\cal C}$, ${\cal C}$, $B$ version of Jacobi's identity.
This {\sl decouples} finding canonical observables to be  
{\sl after} establishing Constraint Closure of the constraints provided by Temporal and Configurational Relationalism (c.f.\ Fig 1.b).  
Each closed subalgebraic structure of constraints moreover furnishes its own notion of canonical observables.
The most salient two cases are no constraints, returning the unrestricted functions  $U$              (if suitably smooth),      and all first-class constraints, 
                                               returning the {\it Dirac observables} $D$ \cite{DiracObs} (the fully physical case).
Some physical theories support further intermediate notions, e.g.\ Kucha\v{r} observables $K$  \cite{Kuchar93} for theories in which the first-class linear constraints close 
[including GR, since ${\cal M}_i$ closes by (\ref{Mom,Mom})]. 
More generally, each physical theory's constraint algebraic structure supports a bounded lattice of subalgebraic structures, 
supports in turn a {\sl dual bounded lattice of notions of observables} \cite{AObs3}. 
$U$ and $D$ notions are supported by {\sl all} theories as the top and bottom elements of the bounded lattice, 
whereas other notions, including the $K$, are not universal theory-independent notions.  
Relational Mechanics supports both Kucha\v{r} observables and `chronos observables' commuting with its case of ${\cal C}\mh\mr\mo\mn\mo\ms$: ${\cal E}$.

\m

\n Finally, as regards not only defining suitable concepts of observables but also finding a full complement of observables 
(each type itself forms an algebraic structure, by the $B, B, {\cal C}$ Jacobi identity), 
\cite{AObs2, ABook} further formulate the bracket conditions (\ref{CO}) as specific DEs, for which \cite{AObs4} provides further specific methodology (for now for finite cases).

\section{Spacetime Construction, Relationalism and Refoliation}

\n The path through the facets branches after the first three, as per Fig 1.b).   
The remaining fork starts by taking {\sl a family of} candidate theories through the (differential almost) Dirac Algorithm. 
E.g.\ 

\n \be
S^{w,y,a,b}  \es  \iint \d^3x \sqrt{\sqrt{h}\{a \, R + b\}} \, \d s_{w,y}  \m ,
\label{trial} 
\ee
where $\d s_{w,y}$ is built out of the usual $\d - \pounds_{\d\u{F}}$ and the more general if still ultralocal supermetric 

\n $M^{abcd}_{w,y} := \sqrt{h}\{h^{ac}h^{bd} - w \, h^{ab}h^{cd}\}/y$.  
Its inverse's components are $N_{abcd}^{x,y} := y\{  h_{ac}h_{bd} -  x \,h_{ab}h_{cd}/2 \}/\sqrt{h}$ for  

\n $x := 2w/\{3 w - 1\}$.  
The parametrization by $x$ is chosen for GR to be the $w = x = y = 1$ case. 
%
%
The quadratic primary constraint is now 

\n \be
{\cal H}_{x,y,a,b}  \:=  N_{abcd}^{x,y}p^{ab}p^{cd} - \sqrt{h} \{ a \, R + b \}  
                    \es  0                                                            \m ,
\label{H-trial}
\ee
This forms the Poisson bracket (for $D_i$ the spatial covariant derivative)

\n \be 
\mbox{\bf \{}           (    {\cal H}_{x,y,a,b}    \, | \,    \d \mJ    ) \mbox{\bf , } \, (    {\cal H}_{x,y,a,b}    \, | \,    \d \mK    ) \mbox{\bf \}}  
\es  a \, y \,          (    {\cal M}_i                \, | \,    \d \mJ \, \overleftrightarrow{\pa}^i \d \mK    ) 
\m + \m  2 \, a \, y \{1 - x\} (    D_i p \, | \, {\d \mJ} \, \overleftrightarrow{\pa}^i {\d \mK}   )                                                            \m .
\label{Htrial-Htrial}
\ee
Relative to (\ref{Ham,Ham}), this picks up an {\it obstruction term to having a brackets algebraic structure}. 
This has four factors, each providing a different way in which to attain consistency. 
GR follows from the third of these setting $x = 1$ (see e.g.\ \cite{AM13, ABook} for the other three cases).    

\m 

\n When working with minimally-coupled matter terms, the above obstruction term is accompanied by a second obstruction term.
Its factors give the ambiguity Einstein faced of whether the universal relativity is locally Lorentzian or Galilean (Carrollian featuring as a third option). 
This is now moreover realized in the {\sl explicit mathematical form} of a string of numerical factors 
in what would elsewise be an {\sl obstruction term to having a brackets algebraic structure of constraints}.  
In particular, the GR spacetime solution to the first obstruction term is now accompanied by the condition 
from the second obstruction term that {\sl the locally-Lorentzian relativity of SR is obligatory}.  
This can be viewed as all minimally-coupled matter sharing the same null cone because each matter field is separately obliged to share {\sl Gravity}'s null cone.  

\m

\n Having constructed GR spacetime $\bFrM$, one needs to take into account that it has a relationalism of its own: Spacetime Relationalism.
This is with respect to $Diff(\bFrM)$: unsplit spacetime 4-diffeomorphisms. 
This has its own version of closure -- now not Constraint Closure but Generator Closure, with respect to Lie brackets -- and of observables: 
the {\it spacetime observables} $S$ which Lie-brackets-commute with $Diff(\bFrM)$'s generators.  

\m 

\n We finally address Foliation Independence.  
We formulate the corresponding foliation kinematics along the lines of Isham \cite{I93}. 
Though again a new TRi version (rather than the original ADM-like version) 
is required to retain compatibility with Temporal Relationalism: {\it TRiFol} \cite{TRiFol} (`Fol' standing for foliations). 
This scheme moreover has Refoliation Invariance guaranteed, by a TRi version of {\it Teitelboim's approach} \cite{Tei73}: 
noting that the Dirac algebroid of GR (\ref{Mom,Mom}, \ref{Ham,Mom}, \ref{Ham,Ham}) is none other than a local algebraic formulation of Refoliation Invariance. 
Conversely, resolving Foliation Independence by Refoliation Invariance {\sl requires} an algebroid to keep track of all the foliations.

\section{Quantum Version}

\n I proceed via giving a {\it TRiCQT} (canonical quantum theory) in \cite{ABook}. 
Along the lines of \cite{I84}, this begins with kinematical quantization, and is followed by consideration of quantum constraints $\widehat{\cal C}$ and dynamical quantization.  

\n\be 
\widehat{\cal H}\Psi = 0 
\ee 
is ensuing the {\it Wheeler--DeWitt equation} for the {\it wavefunction of the Universe}, $\Psi$. 

\m 

\n Most of the subsequent work offered is for Semiclassical Quantum Gravity, for which 

\n\be 
\Psi  \es  \me^{i \, S(h)}|\chi(h, \, l)\rangle   \m ,
\ee 
for $h$ slow heavy and $l$ fast light degrees of freedom.
This suffices from the moderate standpoint that this is at most what shall be observed in coming decades. 
The fairly well-known notion of emergent semiclassical time $t^{\sss\se\sm}$ \cite{I93} additionally admits a Machian interpretation \cite{ABook}. 
This needs to be `found afresh' at the quantum level, rather than continuing to use the classical $t^{\se\sm}$, 
for the furtherly Machian reason that abstracting GLET from STLRC is sensitive to partial or total replacement of classical change by {\sl quantum change}, 
e.g.\ of $\d l$ by $\d|\chi(h, \, l)\rangle$.   

\m

\n \cite{ABook} also argues for quantum observables to be `found afresh'. 
This is as opposed to trying to promote classical observables throught the change from Poisson brackets to commutators, 
the restrictions already imposed by kinematical quantization {\sl and} the promotion of classical constraints ${\cal C}$ to their quantized forms $\widehat{\cal C}$.  
This quantum version of our local Problem of Time resolution remains in accord with Fig 1.b)'s order of addressing facets, 
by which the classical version conceptually and technically pre-empts the quantum version.

\section{Conclusion}

\n The approaches outlined in the current Article to each of Fig 1's 7 local facets of the Problem of Time can moreover be pieced together \cite{ABook} to give 
a joint local resolution of the Problem of Time. 
The literature \cite{Kuchar92, I93, APoT, APoT2, ABook} bears witness that previous strategies for individual facets all failed to combine into resolutions of multiple facets at once.
By this, {\sl combining strategies for individual facets is in fact the lion's share of resolving the Problem of Time}. 
This is achieved in \cite{ABook} by developing the strategies indicated in the current Article {\sl well beyond} the remit of previous such strategies, 
as is most manifest by all of TRiPoD, TriFol, TRiCQT... being required in place of just Temporally Relational actions.
\cite{ABook} is thus not only a comprehensive review of the Problem of Time, but also a local resolution thereof.  
This lays out the resolution for Relational Mechanics, Minisuperspace and inhomogeneous perturbations thereabout, as well as for GR. 
The conceptualization and strategies used moreover exhibit considerable universality beyond these examples \cite{ABook}.

\m 

\n Principal remaining frontiers include attempting a global version, topological-level Background Independence, 
and dealing with the possibility of multiple inequivalent quantizations (the so-called Multiple-Choice Problems).   
For these and other frontiers, including over 120 research problems from paper- to program-sized, see \cite{ABook}.  

\m 

\n{\bf Acknowledgments} I thank Chris Isham and Don Page for over a decade of discussions, Julian Barbour for discussions over a decade ago, 
and Enrique Alvarez, Jeremy Butterfield, Anne Franzen, Jonathan Halliwell, Marc Lachieze-Rey, Malcolm MacCallum, Przemyslaw Malkiewicz, Reza Tavakol 
for hosting, support, further discussions.


\end{document}